\documentstyle[epsf, masaps]{revtex}

\def\<{\langle}
\def\>{\rangle}

\date{\today}
\twocolumn
\begin{document}

\title{ New Physics of the $30^\circ$ Partial Dislocation in Silicon
	Revealed through {\em Ab Initio} Calculation}

\author{G\'abor Cs\'anyi$^\dag$, Torkel D. Engeness$^\dag$, Sohrab
	Ismail-Beigi$^\ddag$ and T. A.  Arias$^*$}

\address{$^\dag$Department of Physics, Massachusetts Institute of
	Technology, Cambridge, MA\\$^\ddag$Lawrence Berkeley National
	Laboratory, CA\\ $^*$ Laboratory of Atomic and Solid State
	Physics, Cornell University, NY}

\date{\today}

\abstract{
Based on {\em ab initio} calculation, we propose a new structure for
the fundamental excitation of the reconstructed 30$^\circ$ partial
dislocation in silicon.  This soliton has a rare structure involving a
five-fold coordinated atom near the dislocation core.  The unique
electronic structure of this defect is consistent with the electron
spin resonance signature of the hitherto enigmatic thermally stable R
center of plastically deformed silicon.  We present the first {\em ab
initio} determination of the free energy of the soliton, which is also
in agreement with the experimental observation.  This identification
suggests the possibility of an experimental determination of the
density of solitons, a key defect in understanding the plastic flow of
the material.
\\
\\
\qquad Accepted for publication in the Journal of Physics}
\maketitle\narrowtext

Dislocations are central to the understanding of the mechanical
response of materials.  The mechanical behaviour of any crystalline
material is determined by a hierarchy of crystalline defects of
successively lower dimension.  Grain boundaries are two dimensional
defects that control the evolution of the microstructure of the
material.  The creation and motion of dislocations, which are one
dimensional extended topological defects of the lattice, mediate the
plastic response of a crystal to external stress.  In silicon, which
has a bipartite lattice, the primary mobile dislocations are the screw
and the $60^\circ$ dislocation, which belong to the glide set. They
dissociate into partial dislocations bounding stacking faults.  The
mobility of dislocations in high Peierls barrier materials such as
silicon is effected by the motion of kinks, zero dimensional defects.
The $30^\circ$ degree partial is less mobile than the $90^\circ$
degree partial, so the motion both types of dislocations in the glide
set is controlled by the $30^\circ$ degree partial.  The low energy
kinks along the $30^\circ$ partial have been shown to also involve a
composite structure where kinks bind with soliton excitations in the
reconstructed ground state of the dislocation core.  The solitons
arise because the reconstruction involves pairing of the core atoms,
which leads to two degenerate ground state configurations.  The
domains of the two configurations are separated by the soliton.  These
soliton excitations are also known as ``anti-phase defects''
(APDs)\cite{Bulatov} or ``phase switching defects''. For a general
review on dislocations in semiconductors, see
\cite{Alexander1,Alexander2}, or more recently \cite{KK,JWeber}.

Here we report the results of an {\em ab initio} study exploring the
lattice and electronic structures, excitation energy, and the density
of these APDs which are the simplest, lowest energy, fundamental
excitations of the dislocations in the hierarchy ultimately leading to
the macroscopic mechanical behaviour of the solid.  Using a special
multiscale sampling technique, we are able to present, for the first
time, an {\em ab initio} calculation for the free energy of formation
for this defect.  The soliton is associated with an atom in the
dislocation core which is not part of a reconstructed dimer.  In the
simple, conventional picture, this atom (henceforth to be referred to
as the ``soliton atom'') only has three bonds and therefore an
unpaired electron.  This simple model, however, does not lead to
predictions consistent with of any of the observed ESR signals
associated with plastically deformed silicon.

We propose a new theory for the structure of the
soliton.  We find that the ground state of the soliton has an
unexpected structure with electronic states which {\em are} consistent
with the most stable ESR center in plastically deformed silicon, the
only one which remains after careful annealing.  The reason why the
natural connection between this center and the lowest energy
excitation of the dislocation core has not been made previously is
that the observed ESR center has a highly unusual symmetry.  In
support of our theory for the structure of the soliton, we gather here
several pieces of evidence from both reports of ESR results and our
own {\em ab initio} calculations.  The final combined {\em ab
initio}--experimental identification which we make allows for the
possibility that future more precise measurments of the ESR signal
strength could be used to determine experimentally the soliton density
in plastically deformed silicon, a parameter which is important for
understanding the plastic response of the material.

\begin{figure}
\epsfxsize=3in
\epsffile{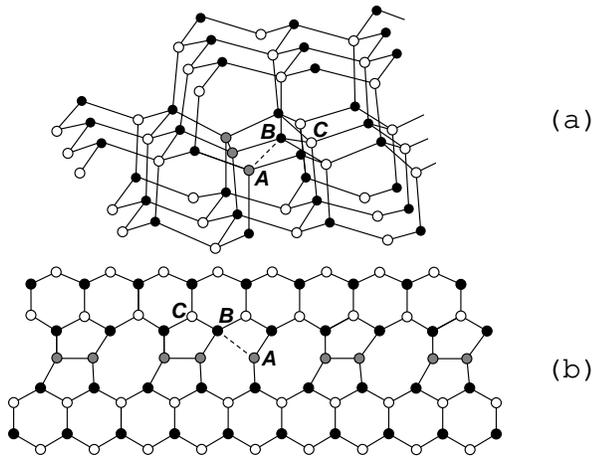}
\smallskip
\caption{(a) Three dimensional view of a $30^\circ$ partial
dislocation, with a soliton domain wall, (b) schematic drawing of the
same in the (111) plane. The grey circles represent the atoms in the
central core of the dislocation.}
\label{fig:disloc}
\end{figure}

Figure \ref{fig:disloc} reviews the basic geometry of the $30^\circ$
partial dislocation studied in this work.  The dislocation is the
one-dimensional boundary defining the edge of a half-planar (111)
stacking fault.  Atoms in the central core of the dislocation (shaded
gray in the figure) are connected to the bulk with only three bonds
per atom.  The dislocation undergoes a reconstruction whereby the core
atoms pair up in dimers forming intra-core bonds and thus become
four-fold coordinated.  This reconstruction breaks the original
translational symmetry along the core and doubles the primitive repeat
distance along the core axis.  Associated with this broken
translational symmetry is a low energy soliton defect where, by
creating a single unpaired core atom, the system may change along the
line from one of the two symmetry related degenerate ground state
phases to the other.  Because the soliton atom is expected to have a
dangling bond, it is natural to look for an ESR signal for this
defect.  The relatatively low energy we expect for such an excitation
leads us to expect a relatively large equilibirum population at
silicon annealing temperatures ($\sim 900$K) and therefore that the
ESR signal would not anneal out as quickly as other signals associated
with the formation and motion of dislocations.  We further would
expect this signal to be detected in all systems which contain
$30^\circ$ partial dislocations.

Indeed, it was discovered over thirty years ago that plastically
deformed silicon gives a wide variety of ESR signals
\cite{WeberAlexander,Wohler,EWeber,Suezawa,Kveder,Sieverts}.  Out of
the dozens of ESR centers, four have been identified as associated
with the $30^\circ$ dislocation core.  The literature refers to these
centers as Si-K1, Si-K2, Si-Y and Si-R.  The K2 and K1 defects have
been identified to be electronic excitations of the same structural
defect.  Kisielowski-Kemmerich \cite{KK} made the currently accepted
identification of the K and Y defects.  It is well known that the
first three of the aforementioned ESR signals anneal out over the
(temperature dependent) time-scale of about an hour \cite{Alexander1}.
Only one signal remains, the one labelled R \cite{Kveder,Y-R,Omling}.  This
center is ``thermally stable'' (it does not anneal out) and is
observed even at high deformation temperatures ($> 900^\circ$ K) where
the other signals anneal out too quickly to be observed.  The R center
was shown to be very similar to Y in its near isotropy and large width
and also similar to the ESR signals obtained from amorphous silicon
\cite{Y-R}.  Kisielowski--Kemmerich {\em et al.}  \cite{Y-R} mentioned
the possibility that R is the residue of Y after annealing.  However,
there is no direct evidence that Y center and the R center are due
to the same structural defect.  The intensities of both centers are
proportional to the dislocation density, but the Y intensity is
also proportional to the area swept out by the glide of the
dislocations during the deformation. 

The nature of the ESR centers and the soliton excitation energy of the
30$^\circ$ partial dislocation have yet to be addressed with modern
{\em ab initio} techniques.  The APD was studied previously by Heggie
and Jones \cite{Heggie1,Heggie2,Heggie3}. Nunes {\em et al.}
\cite{Nunes} reported results of a tight binding study including the
energetics of kinks and solitons for the $90^\circ$ partial
dislocation.  More recently, Bennetto {\em et al.} \cite{Bennetto}
proposed a new kind of period--doubling reconstruction for the
$90^\circ$ partial using calculations based on density functional
theory.  Spence {\em et al.}  \cite{Spence} investigated kink motion.
Bulatov {\em et al.}  \cite{Bulatov} have carried out a comprehensive
study of the $30^\circ$ partial dislocation and its defects but using
only the Stillinger--Weber (SW) inter--atomic potential\cite{SW},
which provides no electronic structure information.  In order to
investigate the electronic structure of low energy excitations of the
$30^\circ$ partial dislocation core, we embarked upon a density
functional study of the system.

To prepare approximately relaxed initial ionic configurations with
the correct bonding topology, we first relaxed lattices containing
dislocation cores using the SW potential.  While doing this, we
discovered that the soliton atom moves out of line with respect to the
dislocation core.  To probe this interesting feature further, we
carried out calculations within the plane wave total energy density
functional approach\cite{RMP}.  To describe the electron--electron
interactions we used the Perdew--Zunger\cite{PerdewZunger}
parameterization of the Ceperly--Alder\cite{CeperlyAlder}
exchange--correlation energy of the uniform electron gas.  To describe
the electron--ion interactions we used a non--local pseudopotential of
the Kleinmann--Bylander form\cite{KB}.  The electronic wave functions
were expanded in a plane-wave basis up to a cutoff of $8$ Ry.

All super-cells used in this study have the same size in the plane
perpendicular to the (110) dislocation axis.  Two partial dislocations
of equal but opposite Burgers vectors at a separation of $14$ \AA\ cut
through this plane.  Following Bigger {\em et al.} \cite{Bigger}, the
lattice vectors are arranged so that the periodic dislocation array
has a quadrupolar arrangement.  Each cell contains forty-eight atoms
per core atom in the dislocation core.  To calculate the excitation
energy of the soliton, it is also necessary to calculate the energy of
the perfectly reconstructed dislocation.  However, the smallest
super-cell which is commensurate with both structures contains six
bilayers ($288$ atoms).  It is possible, however, to reduce the
computational time by using two different super-cells.  For the
reconstructed case, the super-cell contains two bilayers along the
dislocation line, while the soliton structure contains three, (in
total, $96$ and $144$ atoms, respectively).  The lattice vectors were
obtained by relaxing a completely reconstructed dislocation within the
SW model in the $96$ atom cell.  The three bilayer cell was then
obtained from this by multiplying the lattice vector that points along
the dislocation axis by ${\scriptstyle 3\over2}$.  To minimize
numerical errors, including those from {\em k}-point sampling, basis
set truncation and super-cell effects, we compute differences of
energy differences.  For each supercell, we generated a completely
unreconstructed configuration where all the core atoms have only three
bonds.  These structures can be realized in both supercells, and thus
serve as the reference point. The final excitation energy is the
difference between the deviations in the energy from the
unreconstructed structure in each cell.  By keeping the lattice
vectors fixed throughout the calculations, we simulate the strain
field which widely separated solitons would experience along a
reconstructed dislocation line.

To ensure maximum transferabilty of results between the two cells, the
calculations employed k-point sets which give identical sampling of
the Brillouin zone for the two super-cells: $\{(0, 0, \pm 1/4)\}$ for
the $144$ atom cell and $\{(0, 0, \pm 1/6), (0, 0, 1/2)\}$ for the
$96$ atom cell.  To find relaxed structures, we moved the ions
along the Hellmann-- Feynman forces until the ionic forces were less
than $0.02$ eV/\AA.  Typically, this was accomplished in $40$ ionic
steps, where between ionic steps we made $10$--$15$ electronic
relaxation steps using the analytically continued functional
approach\cite{ACprl}.

\begin{figure}
\epsfxsize=3in
\epsffile{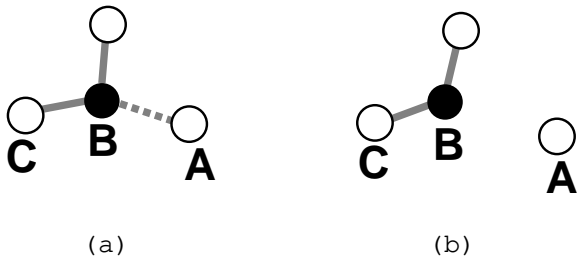}
\caption{Coplanar atoms near the soliton: (a) in the proposed ground
state for the soliton, (b) conventional structure with the soliton
atom in line with the remaining atoms in the dislocation core.  The
notation for the atoms is the same as in Fig. \ref{fig:disloc}: atom A
is the soliton atom, B is the five-fold coordinated bulk atom (see
text), C is a bulk atom bonded to B opposite from A.}
\label{fig:shift}
\end{figure}

Figure \ref{fig:shift} shows the projection in the $(110)$ plane of
our {\em ab initio} results for the structure of the soliton.  The
structure on the left is our prediction for the ground state.  In this
configuration, there is a five-fold coordinated bulk atom (B) in the
immediate neighbouring row to the dislocation core.  Its new, fifth
neighbour is the soliton atom (A).  Our {\em ab initio} results show
that the conventional structure on the right (generated by keeping the
soliton atom collinear with the dislocation core) is not only higher
in energy but also spontaneously decays into the ground state on the
left.

The {\em ab initio} excitation energy of the soliton is
$0.65\pm(\approx0.2$)\ eV, where we attribute most of the uncertainty
to the uncontrolled local density approximation and supercell effects.
If we are to estimate the equilibrium density of the APDs at $900$ K,
the entropy of the system must be taken into account.  Specifically,
we need the change in entropy associated with the creation of an APD.

To determine this, we used a new technique to carry out {\em ab
initio} free energy calculations for the fully reconstructed
dislocation, and the completely unreconstructed dislocation, whose
core can be viewed as a row of APDs.  The entropy change between these
two states gives an estimate of the entropy change of creating a pair
of APDs on the dislocation core.  Traditional techniques for ab initio
evaluation of the entropy is at present infeasable, hence we applied,
for the first time in an {\em ab initio} calculation, the multiscale
sampling approach\cite{Torkel}.  The change in free energy is given by
the adiabatic work of a single degree of freedom in going from the
reconstructed state ($\lambda = 0$) to the unreconstructed state
($\lambda=1$).
$$
\Delta F = \int_0^1 d\lambda \<\partial E/\partial\lambda\>
$$
The essence of multiscale sampling is that the phase space of the
integral is explored in a crude, simple model (in this case,
Stillinger--Weber) using a very large number of samples, resulting in
a much smaller number of statistically independent points.  These are
evaluated within an accurate model ({\em ab~initio}) and a corrective
Boltzmann factor is used to establish the true Boltzmann distribution
of the accurate model.  This method is related to correlated sampling,
or biased sampling.  The key here is that the samples are obtained
from an atomistic, higher level description of the same physical
system, but the exact {\em ab initio} thermal ensemble average is
obtained at a dramatically acccelerated rate (See \cite{Torkel} for
details.)  The sampling was carried out with the standard Monte Carlo
method using the Metropolis algorithm.

\begin{figure}
\epsfxsize=3in
\epsffile{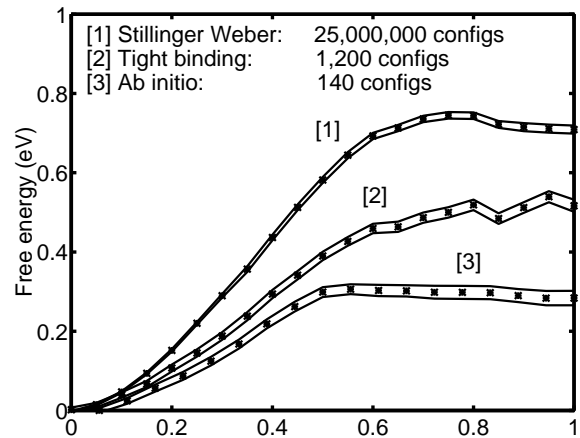}
\smallskip
\caption{Free energy of reconstruction in three different models using
the multiscale sampling approach (see text).  The free energy is
evaulated by integrating the work done in moving the system along the
coordinate $\lambda$ from the reconstructed state to the completely
unreconstructed state.  The lines above and below the datapoints
represent statistical error bars resulting from the sampling.}
\label{fig:freeenergy}
\end{figure}

Figure \ref{fig:freeenergy} shows the free energy as a function of
$\lambda$ for the SW model, a tight binding model and the {\em ab
initio} model, the latter two obtained using multiscale sampling from
the SW model.  Table \ref{tab:freeenergy} shows the entropy calculated
from the free energy.  Although the free energy values are quite
different for the three models (due to the subtleties of the quantum
mechanics of the bond being broken), the entropy variation (mostly
changes in the vibrational modes of the surrounding lattice) is much
smaller.  Taking $T\Delta S$ to be about $0.1$--$0.2$ eV, the free
energy difference associated with creating an APD can be estimated to
be about $0.3$--$0.6$ eV.  This corresponds to a density
$$\rho = {\scriptstyle{1\over2}}e^{-\Delta F/kT}$$ in the range of
$4\times10^{-4}$ to $10^{-2}$ solitons per core atom at $900$ K, which
is consistent with the observed ESR signal strengths; the R center is
estimated to have a density of about $0.001$ per core atom
\cite{Omling,Kimerling}.  (The factor of one half in $\rho$ comes
from the fact that in a particular given phase of the ground state,
only half of the core atoms represent possible soliton sites.)  Given
the exponential sensitivity of the density, we find this agreement
encouraging, with the caveat that the exact temperature at which the
density of solitons is frozen during the quenching after the annealing
is at present unclear.

\begin{figure}
\epsfxsize=3in
\epsffile{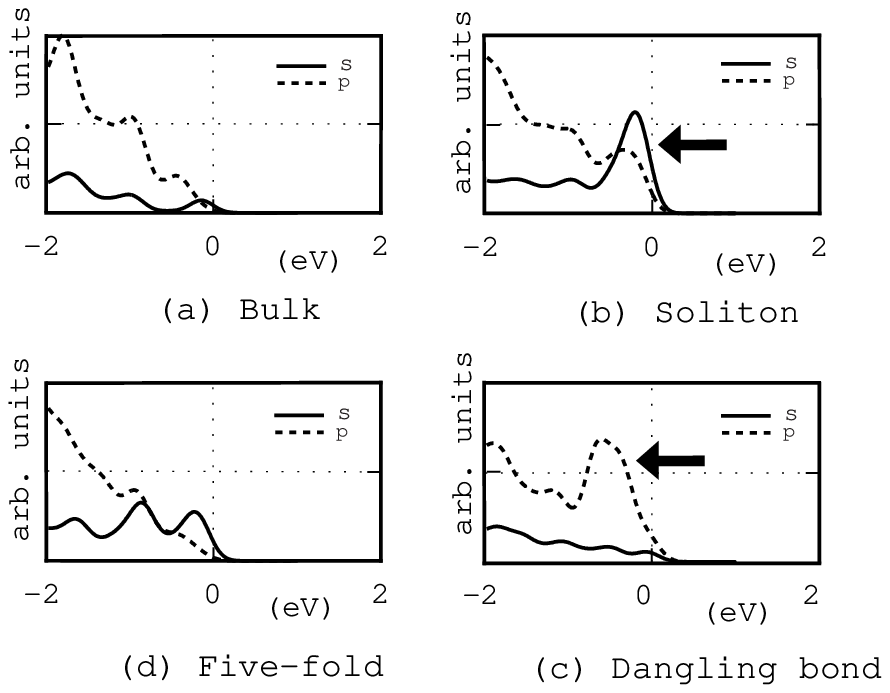}
\smallskip
\caption{Local density of states, calculated by acting on the filled
bands with the Kleinmann--Bylander projectors centered at (a) an atom
deep in the bulk , (b) the soliton atom in its proposed ground state,
(c) a conventional soliton atom with a dangling bond, (d) the quasi
five--fold coordinated atom.  The horizontal axis is the energy (eV),
the scale of the vertical axis is arbitrary but the same for all four
panels.  Solid and dashed lines represent densities in the {\em s} and
{\em p} channels respectively.}
\label{fig:ldos}
\end{figure}

A great advantage of the {\em ab initio} calculations, beyond their
accuracy, is that they also yield the electronic states, in particular
giving information about their spatial symmetry.  In Figure
\ref{fig:ldos} we plot the angular momentum decomposition of the local
density of states as obtained from the Kleinmann--Bylander projections
of the electronic eigenstates.  Part (a) shows, for an atom far from
the core, the familiar concentration of s--like states at the bottom
of the valence band and p--like states at the top.  To explore the
nature of the soliton state, we compare this to the local densities of
states for the soliton atom in the proposed (\ref{fig:ldos}b) and
conventional (\ref{fig:ldos}c) configurations.  We also plot the local
density of states for the quasi-fivefold coordinated atom
(\ref{fig:ldos}d).  The appearance of the peak near the top of the
valence band in the {\em s} channel of the soliton atom in its ground
state (\ref{fig:ldos}b) shows that the state associated with the
soliton is much less anisotropic than the simple dangling p--like bond
on the soliton atom in the conventional picture (\ref{fig:ldos}c).
Note also that the density in the {\em p} channel of the soliton atom
is also correspondingly diminished relative to that in the
conventional state.  We further note an enhancement at the same energy
in the {\em s} channel of the quasi-fivefold coordinated atom
(\ref{fig:ldos}d), which indicates that the unpaired electron is
shared between this atom and the soliton atom.  Defects in the
dislocation core therefore need not be associated with strongly
directional electronic states, as has been previously assumed in
identifications of ESR centers.  Similar mechanisms can plausibly play
a role in the decrease in anisotropy of other point defects.

The connection between the symmetry of the electronic state and the
corresponding ESR signal is via the effective $g$ tensor,
$$
g_{ij} = g_0 \delta_{ij} - 2\Lambda\sum_n\frac{\<\phi_0|{\bf
L}_i|\phi_e\>\<\phi_e|{\bf L}_j|\phi_0\>}{E_e-E_0},
$$
where $\phi_0$ is the unpaired state and $\phi_e$ are the excited
states and $\Lambda$ is the atomic spin-orbit coupling constant.  In
general $\phi_0$ can be broken into angular momentum components (as in
Figure \ref{fig:ldos}), of which the $s$ wave component makes no
contribution to the off-diagonal matrix elements, so we expect the
anisotropy of $g$ to be proportional to the population of the $p$
channel.  (Higher angular momentum components are negligible for
filled states in silicon.)  This population, the area under the peak
associated with the unpaired electron in the {\em p} channel, drops by
about a factor of two as the soliton moves from its symmetrical
dangling bond configuration (\ref{fig:ldos}c) to our proposed state
(\ref{fig:ldos}b).  The literature contains qualitative observations
of the decreased anisotropy of the R signal and one quantitative
comparison which comes from measurements of the Y signal, of which the
R is presumed to be the residual after annealing.  In \cite{KK}, this
anisotropy is compared directly with that of the K1,2 centers, which
have typical dangling bonds, and is shown to be less by about a factor
of two, in agreement with our electronic structure results.

\begin{figure}
\epsfxsize=3in
\epsffile{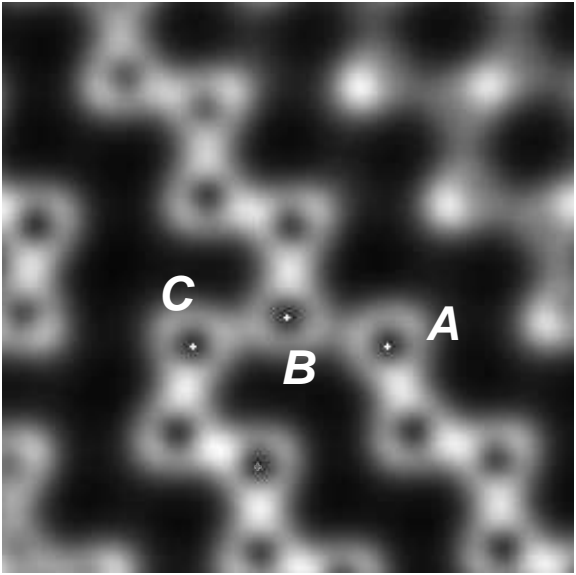}
\smallskip
\caption{Two dimensional slice of the total charge density from our
{\em ab initio} calculation, through the plane containing the atoms A,
B, C, as labelled in the previous figures. The new A--B bond is nearly
as pronounced as the bulk B--C bond, which is clearly weakened
compared to the other (vertical) bulk bond of atom B seen in the
figure.}
\label{fig:dens}
\end{figure}

To explore the nature of the bonding near the soliton atom, we plot
the total valence charge density in Figure \ref{fig:dens}, which shows
that the soliton atom (A) makes a weak bond with the neighbouring
five--fold coordinated atom in the bulk (B).  The new bond (A--B) of
the bulk atom is very similar to a now weakened but previously
existing bond in the spatially opposite direction (B--C). Such
five-fold coordinated structures have been considered previously in
silicon by Pantelides \cite{Pant} and more recently by Duesbery et
al.\cite{Duesbery}.  Through {\em ab initio} studies it was
demonstrated that such five-fold defects in amorphous silicon should
show similar anomalies in the angular momentum decomposition of the
local density of states\cite{floatDFT}.  In amorphous silicon,
however, the shift from the {\em p} channel to the {\em s} channel is
more pronounced with the peak in the {\em p} channel disappearing
completely.

In conclusion, we have presented {\em ab initio} results indicating
that the ground state of the soliton has an unusual structure
involving a five-fold coordinated atom and a correspondingly unusual
electronic structure.  The excitation energy we calculate with the
first {\em ab initio} application of the multiscale approach for this new
defect corresponds to a thermal equilibrium density which is
compatible with the observed signal strength of the R center, which is
the only thermally stable paramagnetic center associated with the
$30^\circ$ partial dislocation.  In line with our notion of the
soliton being the fundamental excitation of the reconstructed
dislocation core, the R center is observed independent of the method
of deformation and in proportion to the dislocation density.  Our
calculations show that the soliton has an enhanced isotropy compared
to that of a simple dangling bond, which correlates well with the
puzzling, nearly isotropic signal of the R center.  In the ground
state structure which we propose, the soliton atom makes a weak bond
with a neighbouring bulk atom and thus gives rise to an amorphous-like
bonding arrangement.  This could explain in part the similarity of the
ESR signature of the R center to that of amorphous silicon.  Based on
the above arguments and results, we propose that the domain walls in
the reconstruction of the $30^\circ$ partial dislocation and the R
centers observed in ESR experiments are one and the same.  Any viable
competing theory which does not identify the R signal with the soliton
must both predict a more plausible microscopic structure for the R
center and explain why the unpaired electron of the low energy soliton
does not exhibit an ESR signal.  

It must reemphasized that the above results and indentification relate
to thermally equilibrated paramagnetic centers.  It is not clear
at present, what is the precise connection between the solitons and
the Y center.  Furthermore, in light of this new proposal for the 
native defect of the system, it might be time to reexamine
the problem of the C line in the DLTS experiments on plastically
deformed silicon.  This, so far unidentified defect is also thermally
stable, so it would be natural to investigate its relationship to 
the soliton of the $30^\circ$ degree partial.

\begin{acknowledgements}
The calculations were carried out on the Xolas prototype SMP cluster.
This work was supported primarily by the MRSEC Program of the National
Science Foundation under award number DMR 94-00334 and also by the
Alfred P. Sloan Foundation (BR-3456).  CSG would like to thank
Vitaly~Kveder, Wolfgang~Schr\"oter and Eicke~Weber for very useful
discussions.
\end{acknowledgements}

\begin{table}
  \begin{tabular}{|cccc|} \hline
  Model & $\Delta$E (eV) & $\Delta$F (eV) & $T \Delta S$ (eV) \\ \hline
  SW & $0.81$ &  $0.71\pm 0.01$ &
  $0.10 \pm 0.01$\\
  Tight binding & $0.71$ & $0.53\pm 0.02$
 & $0.18 \pm 0.02$\\
  {\em Ab initio} &  $0.42$ & $0.27\pm 0.03$ & $0.15 \pm 0.03$ \\ \hline
  \end{tabular}
\caption{Free energy and entropy of reconstruction.  The statistical
uncertainty from the multiscale sampling method is shown.  The
reconstruction energy is displayed in the first column for
information.}
\label{tab:freeenergy}
\end{table}

\end{document}